\documentclass[10pt,prb,twocolumn,showpacs]{revtex4}
\usepackage[dvipdfm]{graphicx}
\usepackage{amsmath,amssymb,multirow}
\newcommand{\be}{\begin{equation}}
\newcommand{\ee}{\end{equation}}
\newcommand{\m}{{\bf m}}

\begin{document}
\title{Full counting statistics of chaotic cavities with many open channels}
\author{Marcel Novaes}
\affiliation{School of Mathematics, University of Bristol, Bristol
BS8 1TW, UK}

\begin{abstract}
Explicit formulas are obtained for all moments and for all cumulants
of the electric current through a quantum chaotic cavity attached to
two ideal leads, thus providing the full counting statistics for
this type of system. The approach is based on random matrix theory,
and is valid in the limit when both leads have many open channels.
For an arbitrary number of open channels we present the third
cumulant and an example of non-linear statistics.
\end{abstract}


\maketitle

The physics of current fluctuations in mesoscopic conductors is an
interesting and fundamental quantum mechanical problem, since at low
temperatures they are mainly due to the discreteness of the electron
charge. The study of shot noise, for example, is an active area of
theoretical and experimental research involving different types of
systems (e.g. quantum dots, disordered wires, quantum point
contacts) and different regimes \cite{pr336ymb2000} (e.g. Coulomb
blockade, quantum Hall effect, localization). A relatively recent
approach is the concept of full counting statistics
\cite{prb51hl1995}, the study of all cumulants of the charge
fluctuation, which amounts to having complete information concerning
charge counting in a transport process. This approach has recently
attracted much attention and has been applied in a wide variety of
situations (see Ref.~\onlinecite{noise} and references therein).
Experimental measurements of the third moment of these fluctuations
have already been reported \cite{prl91br2003}.

In the case of chaotic cavities the random matrix theory (RMT)
approach \cite{rmp69cwjb1997} has been very successful in
reproducing different experimental observations related to quantum
transport, such as weak localization and universal conductance
fluctuations. If the cavity is connected to two ideal leads,
supporting respectively $N_1$ and $N_2$ open channels, the
conductance is given by the Landauer-B\"uttiker formula $g=G_0{\rm
Tr}[T]$, where $G_0$ is the conductance quantum, $T=t^\dag t$ and
$t$ is the transmission matrix. However, only very recently was an
exact expression obtained \cite{prb73dvs2006} within RMT for the
shot noise $P=P_0{\rm Tr}[T(1-T)]$ (with $P_0=2eVG_0$ where $V$ is a
small voltage bias) and an explicit general result is not available
for higher moments of the type ${\rm Tr}[T^m]$. For chaotic cavities
with large $N_1, N_2$ the third and fourth cumulants of charge
transfer have been obtained \cite{pe11ymb2001}, as well as an
expression for the cumulant generating function \cite{jsm2005}.

On the other hand, recent semiclassical calculations based on
correlated classical trajectories that transmit through the cavity
have been able to reproduce the RMT results, both for the
conductance and for the shot noise. These calculations have a
natural perturbative structure on the parameter $N^{-1}$, where
$N=N_1+N_2$ is the total number of channels. Initially the leading
order expressions were reproduced, \cite{prl89kr2002,prl96rsw} later
the full series were obtained and exactly summed. \cite{prl96sh2006}
The next natural step would be to tackle ${\rm Tr}[T^m]$, and it is
thus of interest to have the corresponding RMT prediction of this
quantity, at least to leading-order in $N^{-1}$. This is the purpose
of this work.

We will be interested in the dimensionless moments defined as
$\sum_{i=1}^n T_i^m$, where $T_i$ are the eigenvalues of the matrix
$T$ and $n={\rm min}\{N_1,N_2\}$. Within RMT the $T_i$ are
correlated random numbers between $0$ and $1$, whose distribution
depends only on the symmetries of the system (orthogonal, unitary or
symplectic, labeled by $\beta=1,2$ or $4$ respectively). The average
value of the moments are then given simply by \be M_m=n\langle
T_1^m\rangle.\ee The distribution of transmission eigenvalues can be
characterized by a density, $\rho_\beta(T)$, such that $\langle
T_1^m\rangle=\int_0^1\rho_\beta(T)T^mdT$, or equivalently by a joint
probability distribution $\mathcal{P}_\beta$ such that \be \langle
T_1^m\rangle=\int_0^1dT_1\cdots\int_0^1dT_n~T_1^m\mathcal{P}_\beta(T).\ee

The expression for $\mathcal{P}_\beta(T)$ is \cite{rmp69cwjb1997}
\be \mathcal{P}_\beta(T)=\mathcal{N}_\beta^{-1}|\Delta(T)|^\beta
\prod_{j=1}^nT_j^{\alpha},\ee where $\Delta(T)=\prod_{i<j}(T_i-T_j)$
is the Vandermonde determinant,
$\alpha=\frac{\beta}{2}(|N_2-N_1|+1)-1$ and $\mathcal{N}_\beta$ is a
normalization constant. In Ref.~\onlinecite{prb73dvs2006} the
authors used simple recurrence relations from the theory of
Selberg's integral \cite{Mehta} to obtain an exact result with
arbitrary $N_1,N_2$ for the second moment $M_2$ and for the
shot-noise (second cumulant). Here we follow a similar approach and
compute the third cumulant, sometimes called the skewness. Moreover,
we then proceed to obtain explicit formulas for all moments $M_m$
and for all cumulants, valid to first order in the inverse number of
channels, i.e. in the limit $N_1,N_2\gg 1$.

We must note that in the semiclassical limit of short wavelengths
some noiseless scattering states can be created,\cite{prb67} leading
to a breakdown of the universality implied by RMT
predictions.\cite{prl92,jparev} This phenomenon is governed by the
ratio $\tau_E/\tau_D$ of the quantum Ehrenfest time to the classical
dwell time, and its influence has been investigated on
shot-noise,\cite{nature,prl96rsw} the weak localization effect
\cite{prl95} and conductance fluctuations.\cite{prl92,prb73} Our
results are restricted to the universal regime  $\tau_E/\tau_D\to
0$, when these system-specific corrections are neglected.

Let us consider a certain fixed sequence of $k$ positive integers,
$\m=[m_1,...,m_k]$, and for any subsequence of length $q\le k$ let
us define the function \be P^q_{\m}(T)=\prod_{j=1}^qT_j^{m_j}, \quad
P^0_{\m}(T)=1.\ee We take now $T_kP^k_{\m}(T)\mathcal{P}_\beta(T)$,
derive it with respect to $T_k$ and integrate over all variables to
obtain \be\label{F} F=(\alpha+m_k)\langle
P^k_{\m}(T)\rangle+\beta\sum_{j=2}^n\left\langle
P^k_{\m}(T)\frac{T_k}{T_k-T_j}\right\rangle,\ee where the constant
$F$ is given by \be\label{FC}
F=\int_0^1dT_1\cdots\int_0^1dT_n\frac{d}{dT_k}
\left[T_kP^k_{\m}(T)\mathcal{P}_\beta(T)\right].\ee

We can see that $F$ is actually independent of $m_k$. Hence, we may
equate the r.h.s. of \eqref{F} at different values of this variable
arriving at a recurrence relation. To solve this relation in general
is presently beyond reach, but armed with some patience once can
compute the first moments. This is essentially what was done in
Ref.~\onlinecite{prb73dvs2006}. We take it a bit further and find
the third moment. Instead of writing the lengthy expression that
arises for $M_3$ we present the corresponding cumulant (assuming for
simplicity $N_2\ge N_1$), \be\label{q3}
\frac{Q_3}{Q_2}=-\frac{(N_2-N_1+1-\frac{2}{\beta})(N_2-N_1-1+\frac{2}{\beta})}
{(N-1+\frac{6}{\beta})(N-3+\frac{2}{\beta})},\ee where $Q_2$ is the
average shot noise in units of $P_0$, \be \frac{\langle
P\rangle}{P_0}=Q_2=\frac{N_1N_2(N_1-1+\frac{2}{\beta})(N_2-1+\frac{2}{\beta})}
{(N-1+\frac{2}{\beta})(N-2+\frac{2}{\beta})(N-1+\frac{4}{\beta})}.\ee
The result for $Q_3$ agrees in the limit $N\gg 1$ with the one
presented in Ref.~\onlinecite{pe11ymb2001}.

It is also possible to go beyond linear statistics, and compute
higher correlations as for example
\begin{align}\label{vp}&\frac{n(n-1)\langle
T_1T_2(1-T_1)(1-T_2)\rangle}{Q_2}\nonumber\\&=\frac{(N_1-1)(N_2-1)
(N_1-2+\frac{2}{\beta})(N_2-2+\frac{2}{\beta})}
{(N-3+\frac{2}{\beta})(N-4+\frac{2}{\beta})(N-2+\frac{4}{\beta})},\end{align}
a quantity which would be important to compute the variance of the
shot noise.

To be able to arrive at a more general result, we now introduce the
assumption that both leads contain a large number of open channels,
$N_1,N_2\gg 1$, and thus $n\gg k$. In this case the main
contribution to the summation in \eqref{F} will come from $j>k$. We
can thus approximate $F$ by \be F\approx (\alpha+m_k)\langle
P^k_{\m}(T)\rangle+\beta n\left\langle
P^k_{\m}(T)\frac{T_k}{T_k-T_n}\right\rangle.\ee All the results
obtained from now on should be understood as being valid to first
order in $N^{-1}$. Having said that, we drop the ``$\approx$''
symbol and just write equalities.

We can use the identity
\be\left\langle\frac{T_k^m}{T_k-T_n}\right\rangle=\frac{1}{2}\left
\langle\frac{T_k^m-T_n^m}{T_k-T_n}\right\rangle\ee to simplify our
expression for $F$, \be\label{pre} F=(\alpha+\beta n)\langle
P^k_{\m}(T)\rangle+\beta\frac{n}{2}\left\langle
P^{k-1}_{\m}(T)R_{m_k-1}^{k,n}(T)\right\rangle,\ee where
$R_a^{p,q}(T)$ denotes the symmetric polynomial \be
R_a^{p,q}(T)=\sum_{r=1}^aT_p^{a-r+1}T_q^r,\ee and we have neglected
$m_k$ against $\alpha+\beta n$.

Comparing \eqref{pre} for $m_k$ and $m_k-1$ we get the relations \be
\langle P^{k-1}_{\m}(T)T_k\rangle= A_2\langle
P^{k-1}_{\m}(T)\rangle,\ee for $m_k=1$ and more generally
\begin{align}\label{iter} \langle P^{k}_{\m}(T)\rangle&= \langle
P^{k}_{\m}(T)T_k^{-1}\rangle\nonumber\\&+A_1\left\langle
P^{k-1}_{\m}(T)[R_{m_k-2}^{k,n}(T)-R_{m_k-1}^{k,n}(T)]\right\rangle,\end{align}
for $m_k\ge 2$. In the previous equations $A_1$ and $A_2$ are the
constants \be A_1=\frac{\beta n}{2(\alpha+\beta
n)}=\frac{N_1}{N},\quad A_2=\frac{2\alpha+\beta n}{2(\alpha+\beta
n)}=\frac{N_2}{N}. \ee Not surprisingly, the parameter $\beta$ has
dropped out of the calculation since leading-order results coincide
for all universality classes. Iterating \eqref{iter} $k$ times we
obtain \be\label{mk} \langle P^{k}_{\m}(T)\rangle= A_2\langle
P^{k-1}_{\m}(T)\rangle-A_1\left\langle
P^{k-1}_{\m}(T)R_{m_k-1}^{k,n}(T)\right\rangle.\ee

Since we are interested in moments, we consider a particular case of
the previous equation which is \be\label{eq1} \langle
T_1^m\rangle=A_2-A_1\langle R_{m-1}^{1,2}(T)\rangle.\ee On the other
hand, Eq. \eqref{mk} also gives \be\label{eq2} \langle
R_{m}^{1,2}(T)\rangle=A_2\sum_{j=1}^m\langle
T_1^j\rangle-A_1\sum_{j=1}^{m-1}\left\langle
T_1^{m-j}R_j^{2,3}(T)\right\rangle.\ee We must remark that the
exponents of the terms inside the last brackets provide all ordered
partitions of $m$ into $3$ positive integers. These two equations
can now be iterated together to yield the moments $M_m=n\langle
T_1^m\rangle$, which will in fact be a polynomial of degree $m$, \be
M_m(\xi)=N\sum_{p=1}^m C_{mp}\xi^p, \quad \xi=\frac{N_1N_2}{N^2}.\ee

Finding out the coefficient $C_{mp}$ of the power $\xi^p$ is now an
exercise in combinatorics. The first part of the problem consists in
answering the following question: In how many ways can one build
sequences $\{a_1,...,a_{2p}\}$ with $a_j\in\{A_1,A_2\}$ such that
both $A_1$ and $A_2$ appear exactly $p$ times and in all
subsequences $\{a_1,...,a_{q}\}$, $q<2p$ the number of $A_2$'s is
not larger than the number of $A_1$'s. The solution to this classic
problem are the celebrated {\it Catalan numbers},\cite{combina} \be
c_p=\frac{1}{p+1}\binom{2p}{p}.\ee The power $\xi^p$ in $M_m(\xi)$
will thus contain a factor $(-1)^{p-1}c_{p-1}$. It will also be
multiplied by another factor, which is equal to the number of
ordered partitions of $m$ into $p$ positive integers. This is
$\binom{m-1}{p-1}$.

\begin{table}[t]
\begin{tabular}{|c c| *{7}{r}|}
    \hline
\multicolumn{2}{|c}{$C_{mp}$ }  &  \multicolumn{7}{|c|}{$p$} \\
\multicolumn{2}{|c|}{ }  & 1 & 2 & 3 & 4 & 5 & 6&7\\
    \hline
\multirow{7}{*}{$m$} & 1 & 1 &  & & & &&\\
&2 & 1 & -1 & & & &&\\
&3 & 1 & -2 & 2 & & &&\\
&4 & 1 & -3 & 6 & -5 & &&\\
&5 & 1 & -4 & 12 & -20 & 14 &&\\
&6 & 1 & -5 & 20 & -50 & 70 & -42&\\
&7 & 1 & -6 & 30 & -100 & 210 & -252 & 132\\
    \hline
\end{tabular}
\caption{The values of the moment coefficients $C_{mp}$ for several
values of $m$ and $p$.}
\end{table}

We thus obtain our main result, an explicit expression for all the
moments, valid to first order in the inverse number of channels:
\be\label{mom}
M_m(\xi)=N\sum_{p=1}^m\binom{m-1}{p-1}(-1)^{p-1}c_{p-1}\xi^p.\ee The
first three moments agree with known results.\cite{pe11ymb2001} We
present the coefficients $C_{mp}$ with $m$ up to $7$ in Table I.

The following equation, \be\label{m2q} \sum_{i=1}^n \ln
\{1+T_i[e^{\lambda}-1]\} = \sum_{k=1}^\infty \frac{\lambda^k}{k!}
Q_k, \ee relates the moments and the cumulants.\cite{prb51hl1995}
These will also be given by polynomials, $Q_k(\xi)=N\sum_{p=1}^k
D_{kp}\xi^p$. By feeding \eqref{m2q} with our result \eqref{mom} we
can obtain the first few cumulants, and the coefficients $D_{kp}$
are shown in Table II. We have found by direct inspection that these
coefficients are such that \be\label{cum} Q_k(\xi)=N\sum_{p=1}^k
(-1)^{k+p}\frac{(2p-2)!}{p!}S(k-1,p-1)\xi^p,\ee where \be
S(k,p)=\frac{1}{p!}\sum_{j=0}^p(-1)^{p-j}\binom{p}{j} j^k\ee are the
{\it Stirling numbers} of the second kind.\cite{combina} From the
cumulants we can derive the generating function \be
\sum_{k=1}^\infty \frac{\lambda^k}{k!}Q_k(\xi)=2N\xi\int_0^\lambda
\frac{dz}{1+\sqrt{1+4\xi(e^{-z}-1)}},\ee which is in fact equal to
the one obtained in Ref.~\onlinecite{pe11ymb2001}, thus implying the
correctness of \eqref{cum}. \cite{thanks}

\begin{table}[b]
\begin{tabular}{|c c| *{7}{r}|}
    \hline
\multicolumn{2}{|c}{$D_{kp}$ }  &  \multicolumn{7}{|c|}{$p$} \\
\multicolumn{2}{|c|}{ }  & 1 & 2 & 3 & 4 & 5 & 6&7\\
    \hline
\multirow{7}{*}{$k$} & 1 & 1 &  & & & &&\\
&2 & 0 & 1 & & & &&\\
&3 & 0 & -1 & 4 & & &&\\
&4 & 0 & 1 & -12 & 30 & &&\\
&5 & 0 & -1 & 28 & -180 & 336 &&\\
&6 & 0 & 1 & -60 & 750 & -3360 & 5040&\\
&7 & 0 & -1 & 124 & -2700 & 21840 & -75600 & 95040\\
    \hline
\end{tabular}
\caption{The values of the cumulant coefficients $D_{kp}$ for
several values of $k$ and $p$.}
\end{table}

In summary, we have explicitly obtained the random matrix theory
prediction for all moments and all cumulants of the charge current
in a chaotic cavity, in the limit of large channel numbers.
Naturally, it would be desirable to obtain such explicit expressions
for arbitrary channel numbers, but in this case we were able to
compute only special cases such as \eqref{q3} and \eqref{vp}. The
moments are natural quantities to be studied in semiclassical
approaches to the problem, and indeed Eq. \eqref{mom} has been
reproduced using action-correlated trajectories in the open quantum
star graph \cite{unpub}. The Hamiltonian case and corrections due to
finite Ehrenfest time are discussed to some extent in Ref.
\onlinecite{brouwer}.

Support by EPSRC is gratefully acknowledged.


\begin{thebibliography}{99}

\bibitem{pr336ymb2000} Ya.M. Blanter and M. B\"uttiker, Phys. Rep.
{\bf 336}, 1 (2000).

\bibitem{prb51hl1995} L.S. Levitov, and G.B. Lesovik, JETP Lett. {\bf 58}, 230 (1993);
H. Lee, L.S. Levitov and A.Yu. Yakovets, Phys. Rev. B {\bf 51}, 4079
(1995).

\bibitem{noise} Yu. V. Nazarov (ed), {\it Quantum Noise in Mesoscopic
Physics} (Dordrecht, Kluwer, 2003).

\bibitem{prl91br2003} B. Reulet, J. Senzier and D.E. Prober, Phys.
Rev. Lett. {\bf 91}, 196601 (2003); Yu. Bomze {\it et al}, {\it
ibid} {\bf 95}, 176601 (2005).

\bibitem{rmp69cwjb1997} C.W.J. Beenakker, Rev. Mod. Phys. {\bf 69},
731 (1997).

\bibitem{prb73dvs2006} D.V. Savin and H.-J. Sommers,
Phys. Rev. B {\bf 73}, 081307(R) (2006).

\bibitem{pe11ymb2001} Ya.M. Blanter, H. Schomerus and C.W.J.
Beenakker, Physica E {\bf 11}, 1 (2001).

\bibitem{jsm2005} O.M. Bulashenko, J. Stat. Mech. P08013 (2005).

\bibitem{prl89kr2002} K. Richter and M. Sieber, Phys.
Rev. Lett. {\bf 89}, 206801 (2002); H. Schanz, M. Puhlmann and T.
Geisel, {\it ibid} {\bf 91}, 134101 (2003).

\bibitem{prl96rsw} R.S. Whitney and Ph. Jacquod, Phys. Rev. Lett. {\bf 96}, 206804
(2006).

\bibitem{prl96sh2006} S. Heusler, S. Muller, P. Braun and F. Haake, Phys.
Rev. Lett. {\bf 96}, 066804 (2006); P. Braun, S. Heusler, S. Muller
and F. Haake, J. Phys. A {\bf 39}, L159 (2006).

\bibitem{Mehta} M.L. Mehta, {\it Random Matrices} (Academic Press,
New York, 2004), 3rd edition, Chapter 17.

\bibitem{prb67} P.G. Silvestrov, M.C. Goorden and C.W.J. Beenakker, Phys. Rev. B
{\bf 67}, 241301(R) (2003).

\bibitem{prl92} Ph. Jacquod and E.V. Sukhorukov, Phys.
Rev. Lett. {\bf 92}, 116801 (2004).

\bibitem{jparev} I. Aleiner and A. Larkin, Phys. Rev. B {\bf
54}, 14423 (1996); H. Schomerus and Ph. Jacquod, J. Phys. A {\bf
38}, 10663 (2005).

\bibitem{nature} O. Agam, I. Aleiner and A. Larkin, Phys. Rev.
Lett. {\bf 85}, 3153 (2000); S. Oberholzer, E.V. Sukhorukov and C.
Schonenberger, Nature {\bf 415}, 765 (2002).

\bibitem{prl95} S. Rahav and P.W. Brouwer, Phys.
Rev. Lett. {\bf 95}, 056806 (2005).

\bibitem{prb73} S. Rahav and P.W. Brouwer, Phys.
Rev. B {\bf 73}, 035324 (2006).

\bibitem{combina} J.H. van Lint and R.M. Wilson, {\it A Course in
Combinatorics}, (Cambridge, Cambridge University Press, 2001), 2nd
edition.

\bibitem{thanks} I thank H. Schomerus for pointing this out.

\bibitem{unpub} G. Berkolaiko, J.M. Harrison and M. Novaes,
unpublished.

\bibitem{brouwer} P.W. Brouwer and S. Rahav, cond-mat/0606384.

\end{thebibliography}
\end{document}